\documentclass[]{elsarticle}

\usepackage{hyperref}
\usepackage{amsmath}

\usepackage{caption}
\usepackage{subcaption}
\usepackage{hyperref}
\usepackage{xcolor}
\newcommand{\nn}{\nonumber}
\newcommand{\veps}{\varepsilon}

\begin{document}

\begin{frontmatter}

\title{Bistable colloidal orientation near a charged surface}

\author{Mohit Singh and Yoav Tsori}
\address{Department of Chemical Engineering, Ben-Gurion University of the 
Negev, Israel 8510501}


\begin{abstract}

Anisotropic particles oriented in a specific direction can act as artificial 
atoms and molecules, and their controlled assembly can result in a wide variety 
of ordered structures. Towards this,  we demonstrate the orientation 
transitions of uncharged peanut-shaped	polystyrene colloids, suspended in a 
non-ionic aprotic polar solvent, near a flat surface whose potential is static 
or time-varying. The charged surface is coated with an insulating dielectric 
layer to suppress electric currents. The transition between several orientation 
states such as random, normal or parallel orientation with respect to the 
surface, is examined for two different colloid sizes at low-frequency ($\sim 
10-350$ kHz) or static fields, and at small electric potentials. In  
time-varying (AC) field, a detailed phase diagram in the potential-frequency 
plane indicating the transition between particles parallel or normal to the 
surface is reported. We next present the first study of orientation switching 
in static (DC) fields, where no electro-osmotic or other flow is present. A 
reversible change between the two colloidal states is explained by a theory 
showing that the sum of electrostatic and gravitational energies of the colloid 
is bistable. The number of colloids in each of the two states depends on the 
external potential, particle and solvent permittivities, particle aspect ratio, 
and distance from the electrode.

\end{abstract}

\begin{keyword}
Colloids \sep charged surface \sep bistability \sep electric field \sep
torque
\end{keyword}

\end{frontmatter}


\section{Introduction}

The application of electric fields to suspensions of colloidal particles can 
lead to well-defined structures and ordered phases. Some of the ordered 
structures are potentially useful in electrochemistry 
\cite{ghoroghchian1991gas}, biosensors \citep{velev1999situ,hoettges2003use},  
electronic displays \cite{comiskey1998electrophoretic,hayes2003video}, photonics 
\cite{vlasov2001chip,lumsdon2004two,hosein2010dimer,schuller2010plasmonics} and 
electrical devices \cite{hermanson2001dielectrophoretic,bradley1997creating}. 
Isotropic colloidal particles suspended in a medium can rapidly rearrange upon 
application of an alternating (AC) electric field \cite{lumsdon2004two}. In 
non-conducting fluids, this transition from random to regular arrangement 
depends strongly on the difference between the dielectric constants of 
colloidal particles and the suspending solution. The dielectric contrast induces 
dipole-dipole interactions, which lead to particles chaining one next to the 
other. The colloidal chains orient preferably in the direction of the external 
electric field. In some conditions, the intra-chain interactions render 
two-dimensional hexagonal particle arrays.      
Anisotropic particles can be aligned in electric fields, whether hard or soft, 
charged or neutral. Neutral particles usually align parallel to the electric 
field because the dielectric anisotropy means that this is energetically 
favored. In charged particles, the electric dipole can be due to the migration 
of charges in one direction and the opposite migration of the counterions. 
For example, Yang {\it et al.} \cite{yang2018change} reported the behavior, 
structure, and dynamics of charged asymmetric polystyrene dimers under 
an AC electric field applied perpendicularly to a substrate. They reported 
in-plane assemblies of the charged doublet, triplet, and quadruplet chiral 
clusters with applied low frequencies.

The effect of time-varying electric fields on colloidal orientation is  commonly 
investigated at the high-frequency, MHz, range \cite{dhont2010electric}. At 
such high frequencies, the electric double layer of residual ions in the 
solution vanishes, and electro-osmotic flow is negligible. The structure 
formation at the MHz range can be obtained by applying relatively high electric 
field strengths, which polarize the colloidal core and induce dipole-dipole interactions. 

Much less attention has been paid to low-frequency electric fields, of 
the order of $10$–$100$ kHz. At such low frequencies, several problems are 
encountered, such as abnormal alignment of 
colloids \cite{kramer1994electro}, clumps of rods and spheres 
\cite{mantegazza2005anomalous}, induction in phases, and instability of 
suspensions of long and thin colloidal chains \cite{kang2008double, 
kang2010electric}. In the 
case of anisotropic colloids, short ellipsoidal rods tend to align parallel to 
each other \cite{singh2009one}. However, the polarization of condensed ions, the 
double layer interaction, and electro-osmotic flow may hamper the colloidal 
alignment in low frequencies. Additionally, in such studies, the strength of the 
applied field is much lower than that used in the MHz range.

Recently, a theoretical work showed that an elongated colloid near a charged surface could be aligned 
with its long axis {\it parallel} or {\it perpendicular} to the field. The 
non-classical perpendicular orientation occurs only in spatially non-uniform 
fields. It results from a nontrivial torque that acts on the particle due to the 
presence of an electric double layer. When such a particle is found near a 
charged surface, the colloid orientation can thus switch between two states, 
depending on the surface potential, the distance from the surface, the colloid 
aspect ratio, and other factors \cite{tsori2020bistable}. Unlike the works 
reported in the literature \cite{yang2018change,ma2012two}, we demonstrate 
orientation switching of large uncharged 
peanut-shaped colloids in the presence of small voltage and small or zero-field 
frequencies and negligible currents. A dilute and non-ionic colloidal solution 
is used in the experiments, leading to negligible inter-colloidal 
interaction. The study is focused on the colloids near the charged surface 
and away from colloidal aggregates, and hydrodynamic flow is absent. 
Thus, the forces acting on the colloids are thermal, electrical, and gravity. In 
this first-of-its-kind study, we experimentally demonstrate the orientation 
switching in the simultaneous presence of DC 
electric field and gravity. Experimental observations are explained using a 
unified theory.

\section{Experimental}

\subsection{Materials and methods}

\begin{figure}
\begin{center}
\includegraphics[width=0.9\textwidth,bb=1 405 540 660,clip]{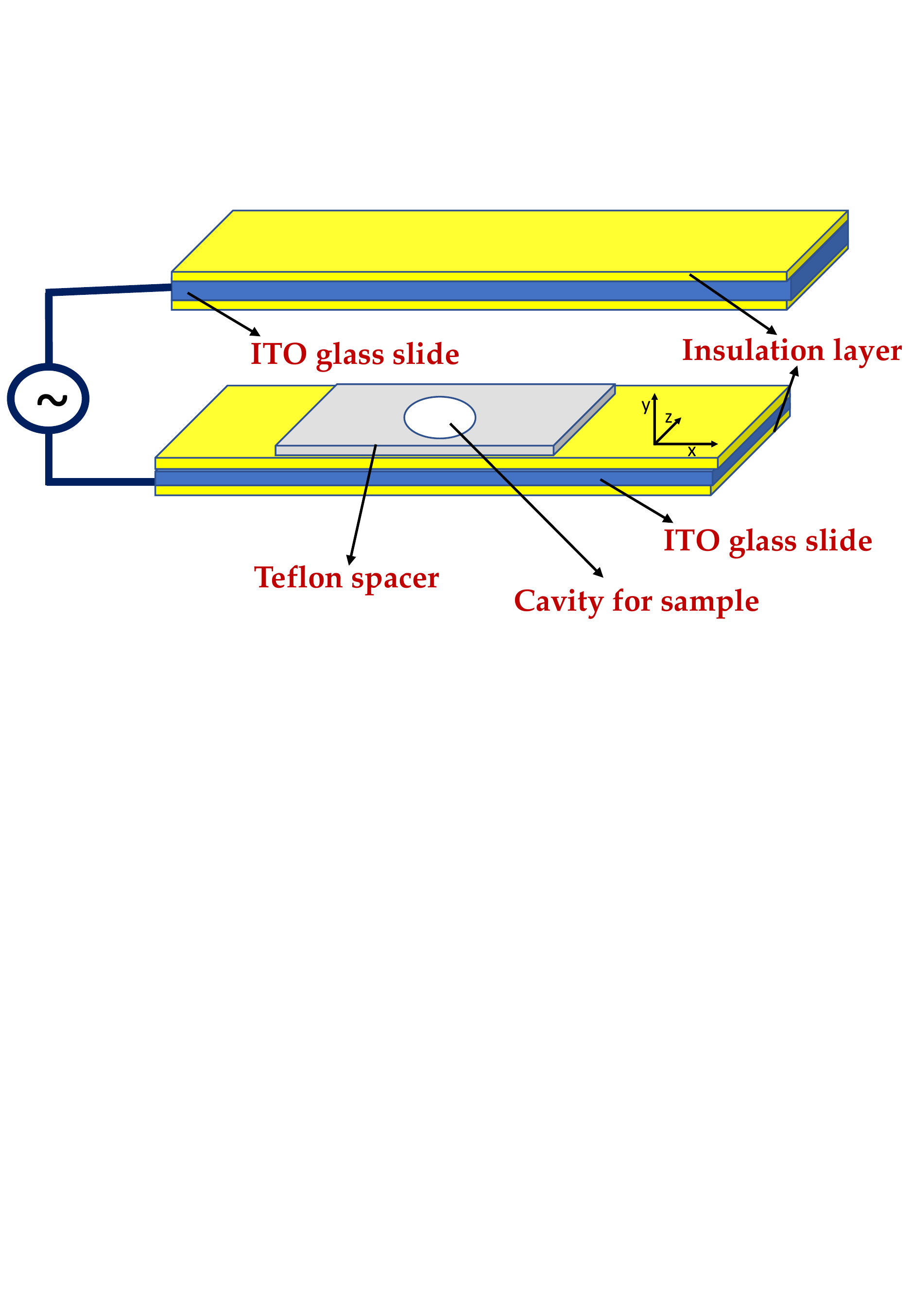}
\caption{Schematic of the experimental setup used to study the colloid 
orientation switching in electric fields. An insulating tape 
(yellow) is used to prevent direct contact between the ITO-covered glass and the 
solution. The sample is observed directly by an optical microscope. }
\label{fig_setup}
\end{center}
\end{figure}
Two 25$\times$75 mm glass slides covered with a thin conducting ITO layer are 
used as electrodes. Electrically-neutral polystyrene peanut-shaped particles 
with reported dimensions of $2.8\times 4.1$ and $5.1\times 7.7$ $\mu$m are 
purchased from Magsphere. Dimethyl sulfoxide (DMSO) (Sigma-Aldrich) is used as a 
solvent. Particle solutions of the two colloid sizes are prepared at a volume 
fraction of $10^{-3}$ by sonicating the solution for $5$ min before each 
experiment. A high-voltage power amplifier (Trek model PZ700-A) with 
voltages up to $700$ V$_{\rm rms}$ and arbitrary waveform generators (OR-X 
function generator 325) are used to generate sinusoidal signals of desired 
voltage. In static field experiments, DC voltage is applied using Keithley 2410 
Source Measuring Unit (SMU) and Trek power amplifier 2210. A camera 
(Thorlabs camera  8050M-CL-TE) fitted onto the top of an inverted phase-contrast 
microscope (Olympus BX 53F) is used to visualize the colloids.

\subsection{Experimental setup}

The experimental setup is shown schematically in Fig. \ref{fig_setup}. Two 
ITO-covered glass slides are placed onto each other; they are separated by the 
non-conductive Teflon tape of thickness $300$ $\mu$m. Similar to the work of 
Kuijk {\it et al.} \cite{kuijk2014effect}, both electrodes are covered with 
$\approx 60$ $\mu$m transparent non-conductive tape to avoid direct contact 
between electrodes and solution. Therefore the field is relatively small, and 
its direction is parallel to gravity. The resulting electrode spacing is 
$\approx 420$ $\mu$m. The suspension is kept inside a circular cavity with a 
diameter of $5$ mm in the Teflon layer. A volume of $30$ $\mu$L solution is 
inserted into the cavity, and the top electrode is gently placed to avoid air 
bubbles. All four corners are clamped using a ``crocodile'' pin clip. The 
ITO-covered glasses are connected via wires to the power supplies. A new device 
is prepared specifically for every experiment to ensure the cleanliness of the 
setup. In the case of AC fields, the ionic conductivity of the solution is below 
the measurement threshold. In experiments with DC potentials in the same setup, 
a fast transient current just after switching on is followed by a decay to zero 
current and a buildup of an electric double layer at the insulating tape.

\subsubsection{Errors in the measurements}

The distribution of colloid orientations is obtained from the microscopy 
images by analysis of the apparent aspect ratios with ImageJ software. At a 
given field of view, not all colloids appear in focus (see inset images in Fig. 
\ref{fig_phase_diagram}); some appear with asymmetrically blurred boundaries. 
This blurriness causes errors in the measurement of the apparent aspect ratio. 
Additionally, for hundreds of colloids in thousands of images, light intensity 
and contrast adjustments may lead to errors in the 
automatic boundary detection process. This results in $\sim 4$\% error in 
the aspect ratio measurements. From Eq. \ref{EQ:AR} (see below) it follows that 
there is an error of $15^\circ$--$20^\circ$ in the calculated tilt angle 
$\theta$.

\section{Results and discussion}

\subsection{Colloidal orientation in AC fields}

In the literature, the colloidal orientation in AC fields is generally 
studied for isotropic particles, with the exception of a few recent works 
\cite{mittal2009electric,mantegazza2005anomalous,demirors2010directed,
hernandez2012role,yan2020single,hendley2021anisotropic,kwaadgras2014orientation,
mohammadimasoudi2016full,crassous2014field,ma2012two}. 
Kuijk {\it et al.} reported a para-nematic phase orientation of nanorods in 
the MHz and $\sim 160$ kHz ranges \cite{kuijk2014effect}. 
Less attention is paid to anisotropic colloids in low-frequency AC fields 
\cite{kuijk2014effect,yang2018change,ma2012two}. Ma {\it et al.}
\cite{ma2012two} investigated the assembly of negatively charged colloidal dimers by 
applying low-frequency AC electric fields and reported that the assembly and 
orientation is a sensitive function of the frequency. They also 
explored the effect of the DC field on the assembly of anisotropic colloids for various 
salt concentrations and found no orientation switching. Unlike the previous 
works, here we use neutral anisotropic colloids and report orientation 
switching in the presence of both AC and DC fields. In the first part of the work, 
we investigate the orientation of anisotropic colloids in low-amplitude and 
low-frequency AC fields as a function of their size, applied frequency, and 
potential. We construct a phase diagram in the frequency-potential plane. 
Polystyrene peanut-shaped prolate ellipsoidal particles with an aspect ratio 
$\sim 1.5$ are dispersed in  DMSO at $1$ wt\%. In the geometry used by us, the 
electric field is approximately uniform throughout the sample. The application 
of alternating fields periodically polarizes the colloids according to the 
Maxwell-Wagner-O’Konski mechanism, which is analogous to polarization due to a 
contrast in dielectric properties but also accounts for the contrast in 
conductivity \cite{morgan2003ac}.

We noticed that low-frequency fields in the range of $7$--$350$ kHz cause 
micron-sized colloids to orient rapidly with their long axis parallel to the 
field. Due to the uniformity of the field, electrophoretic or 
dielectrophoretic motion is not observed, and colloidal orientation is 
independent of the distance from the charged surface. In static fields, distance 
matters, as is explained below. 

The colloids are observed in-situ by phase-contrast microscopy as 
cross-sections in the $x$-$z$ plane at a variable distance $y$ from the 
electrodes (see Fig.\ref{fig_setup}). In the absence of an electric field, the 
colloidal particles are homogeneously distributed, with a slight tendency to 
settle down at the bottom surface due to the slight difference between the 
specific densities of the colloids and solvent. Their orientation is random due 
to their thermal energy. As we show below, as soon as the electric field is 
turned on above the cut-off field amplitude and frequency, the colloids aligned 
along the field's direction, as manifested by their nearly circular 
cross-sections. If the colloid density 
is sufficiently high, a few minutes after the application of the field,  
colloids start to order in chains along the field's direction. The chain 
formation is reversible; colloids lose orientation and are randomly distributed 
within a few seconds after switching off the field. 

We imaged the colloid as projected onto the x-z plane. In this plane, the 
ellipsoid is an ellipse whose apparent aspect ratio $\text{AR}$ is the ratio between 
the projected semi-major and semi-minor axis. To extract the tilt angle $\theta$ 
of the long axis of the ellipsoid with respect to the $y$-axis, the following 
formula is used:
\begin{eqnarray}\label{EQ:AR}
AR=\sqrt{\left(\frac{b}{c}\right)^2\cos^2\theta+\left(\frac{a}{c}\right)^2\sin^2
\theta}.
\end{eqnarray} 
Here $a>b>c$ are the semi-major axes of the prolate ellipsoid. When the angle is 
$\theta=0^\circ$, the apparent aspect ratio is $AR=b/c$, while 
$\theta=90^\circ$ yields $AR=a/c$ (ellipsoid is lying parallel to the surface). 
One can thus calculate $\theta$ by measuring $\text{AR}$ and using the known values of 
$a$, $b$, and $c$.

\begin{figure}[ht!]
\begin{minipage}{1\textwidth}
\begin{center}
\includegraphics[width=0.45\textwidth,bb=0 0 650 
650,clip]{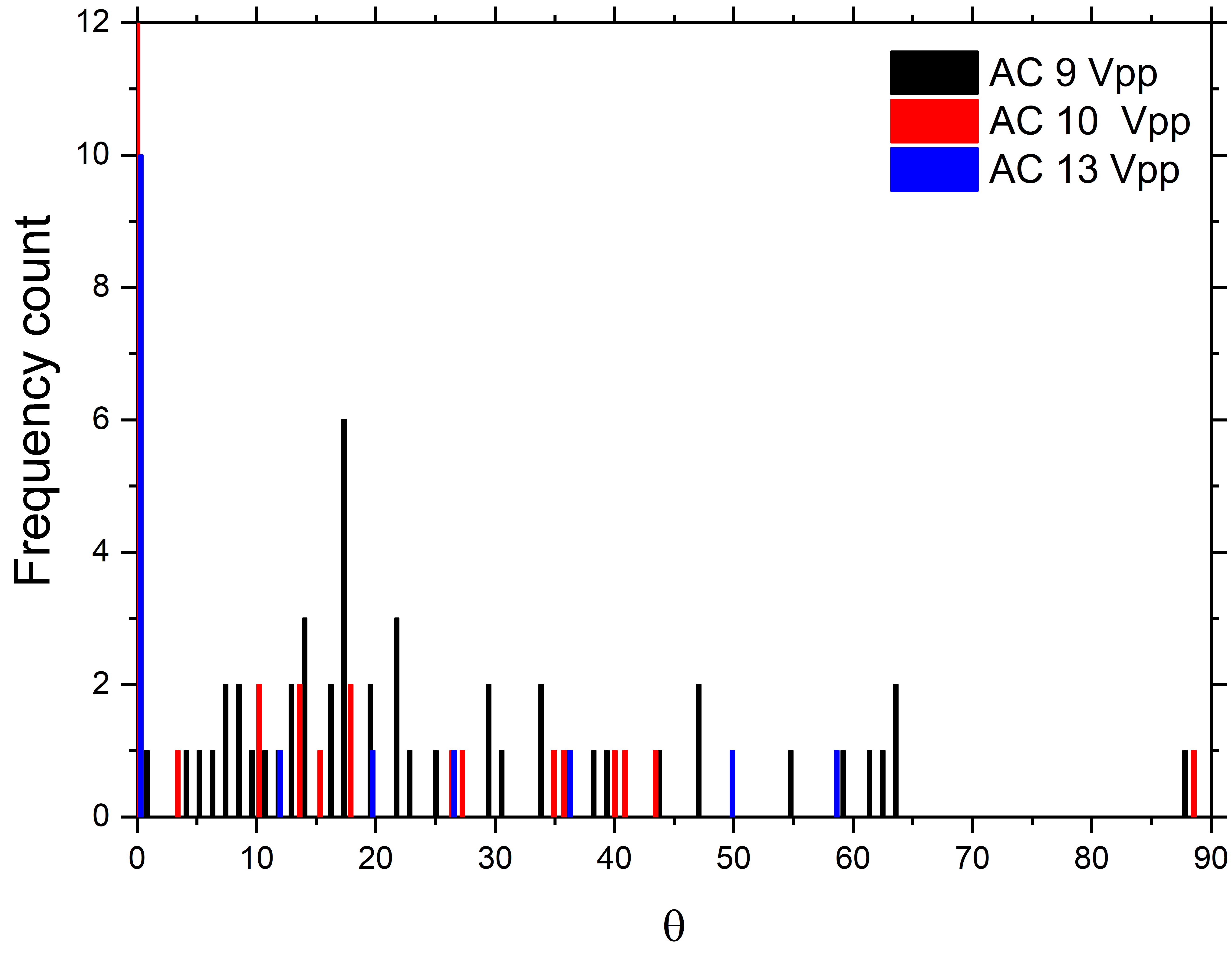}
~
\includegraphics[width=0.45\textwidth,bb=0 0 650 
650,clip]{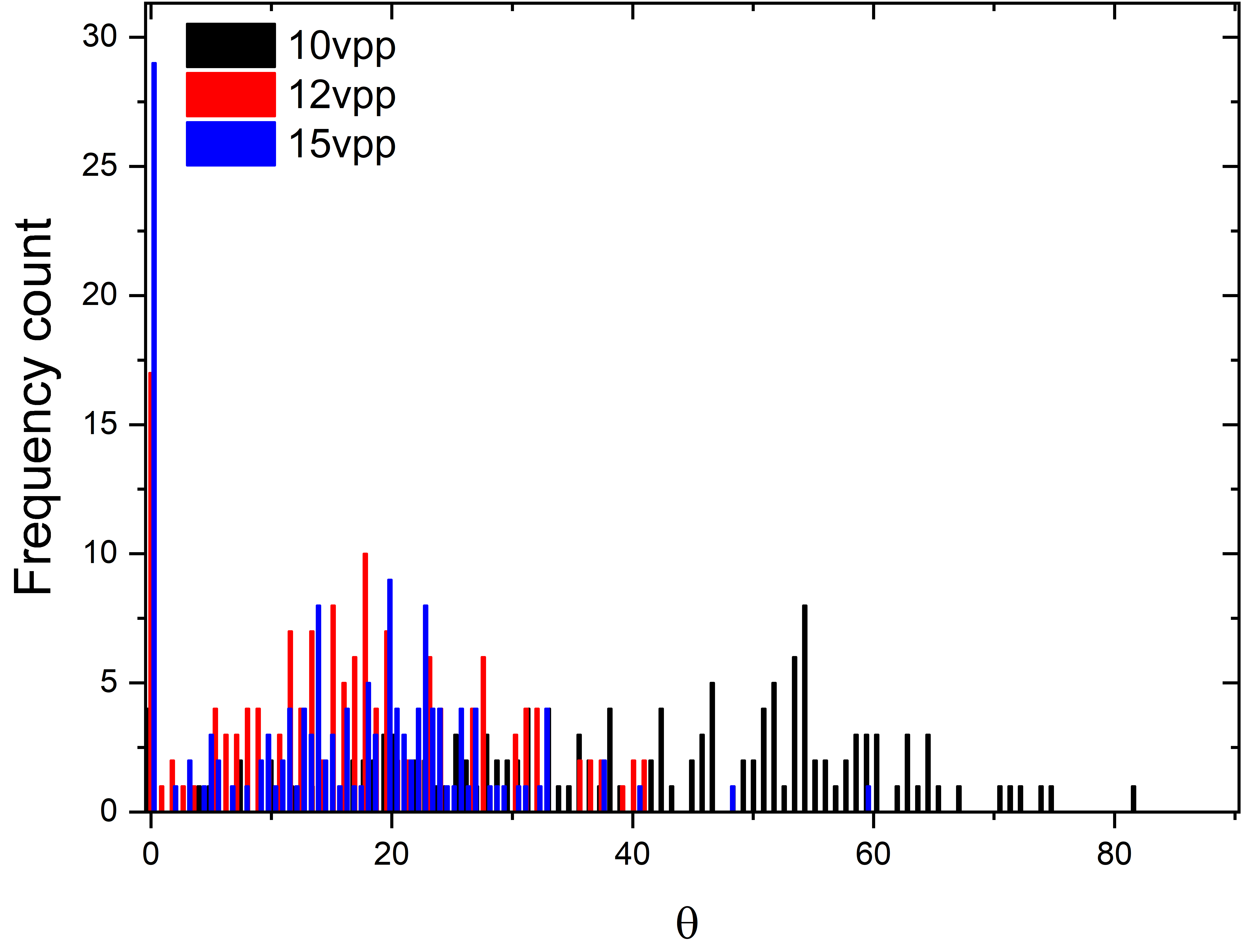}
\end{center}
\end{minipage}
\caption{Distribution of colloid tilt angles $\theta$ with respect to the $y$ 
axis (normal to the electrodes), at fixed frequency and different applied 
voltages. Left: $2.8\times 4.1$ $\mu$m sized colloids at $15$ kHz  
frequency. Right: $5.1\times 7.7$ $\mu$m sized colloids at $25$ 
kHz  frequency.}
\label{fig_AC_voltage}
\end{figure}

The colloid sample is injected between the electrodes, and 
voltage is applied. In Fig. \ref{fig_AC_voltage} we show the measured 
colloidal orientation for two colloid sizes at a fixed frequency and variable 
potentials. In the left panel, the colloids are smaller, and the field 
frequency is $15$ kHz. At the small voltage of $9$ V peak-to-peak (vpp), the 
electrostatic energy is not large enough to overcome thermal fluctuations, and 
the colloidal orientation remained random. At an increased voltage of $10$vpp, 
the orientation shifted toward $\theta=0$, and a more significant number of 
colloids are oriented vertically. Further increase in the voltage to $13$vpp 
resulted in the vertical orientation of a vast majority of the colloids. The 
larger colloids showed a similar phenomenon, as shown in the right panel. At 
$10$vpp, thermal energy overcomes the electrostatic energy, and the colloidal 
orientation is random, with a slight tendency toward $\theta=90^\circ$, due to 
the gravitational force. At $12$vpp, the distribution shifted to smaller values 
of $\theta$, while at $15$vpp, this tendency became even more substantial. More 
significant potentials are required to orient the larger colloids since they are 
more affected by gravity. In these two cases, one would conclude that there are 
two threshold potentials, $10$vpp and $12$vpp, for the small and large 
colloids, respectively. We note that if colloids are treated as dipoles 
lying on a plane, their interaction force is inversely proportional to the 
fourth power of the distance between them (energy varies inversely with 
the cube of the distance) \cite{ma2012two}. The colloidal interaction is 
therefore negligible for the dilute suspension used in the present experiments. 
Additionally, at an applied frequency of the order of $~10^4$ Hz, the electric 
torque is strong enough to align the colloids parallel to the applied field. As 
shown in Fig. \ref{fig_AC_voltage}, the colloidal orientation is random below a 
certain voltage because the electric torque is weak compared to the thermal 
noise.

\begin{figure}[t]
\begin{center}
\includegraphics[width=0.8\textwidth,bb=40 215 480 
545,clip]{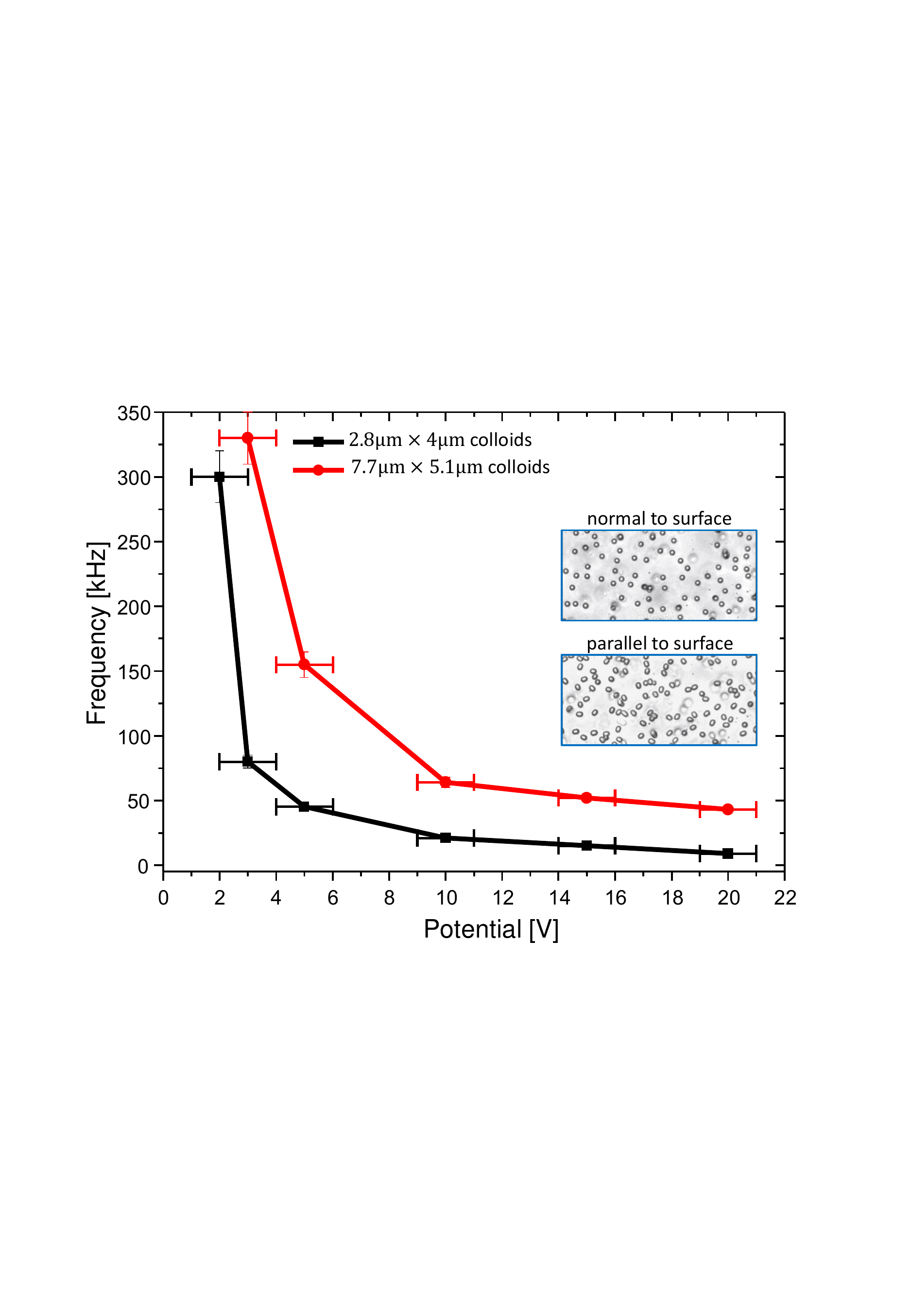}
\caption{Phase diagram of colloidal orientation in AC electric fields. The lines 
signify a transition between two-particle orientations: below the line, 
particles are mainly parallel to the surface, while above it, they are 
perpendicular to it (parallel to the field). The black curve is the threshold 
for small particles, while the red curve is for bigger particles. An increase in 
either potential or frequency leads to particle orientation parallel to the 
field for both sizes. The inset images are representative microscope snapshots 
of the two colloidal states.
}
\label{fig_phase_diagram}
\end{center}
\end{figure}

We set out for a large-scale scan of the 
potential-frequency parameter space to generalize these observations. The 
results are summarized in Fig. \ref{fig_phase_diagram}. The colloidal 
orientation is determined from the peak apparent aspect ratio value at each 
value of potential and frequency. There is a 
line separating perpendicular and parallel orientations in the 
potential-frequency plane. At a fixed potential, an increase of frequency from 
small to large value across this line yields a transition from colloids parallel 
to the surface to perpendicular. The transition frequency is large at small 
potentials, while at high potentials, the transition is at small frequencies. 

The phase diagram has two curves, black and red, corresponding to small and 
large particles, respectively. The larger the colloid, the more this curve is 
displaced upwards and to the right; one needs higher frequencies at given 
potentials or larger potentials at a given field. According to Jones 
\cite{jones1995electromechanics}, anisotropic particles 
align such that the effective energy
\begin{eqnarray}
U=-\frac{1}{2}Re(\alpha_\parallel-\alpha_\perp)E^2 \cos^2(\theta)
\end{eqnarray}
is minimized. Here $\alpha_\parallel$ and $\alpha_\perp$ are the 
effective polarizabilities when particles are aligned parallel or 
perpendicular to the applied field, respectively, and $\theta$ is the angle 
between long axis and the field. The polarizabilities are complex functions of 
the frequency, particle geometry and electrical properties of the solvent and 
particle. For higher aspect ratio and fixed voltage, one has to apply higher 
frequency to keep $\text{Re}(\alpha_\parallel-\alpha_\perp)$ positive. Interestingly, 
and not reported before, the threshold frequency varies non-linearly with 
applied potential. It should be noted that we performed experiments also without 
the insulation layer between the ITO electrode and the solvent.  We found that 
the orientation threshold voltage is greatly reduced to the order of 
millivolts, but it is difficult to determine the transition value accurately. 

\subsection{Colloidal orientation in DC fields}

In this part of the work, we checked the validity of the prediction that in 
constant potentials, colloidal orientation is bistable; namely, they would 
orient perpendicular or parallel to the surface, depending on the surface 
potential \cite{tsori2020bistable}. The ``switching'' behavior depends on the 
colloid size, dielectric constants of the colloid and solvent, and the distance 
from the surface. The driving force for colloidal orientation parallel to the 
surface is the ideal gas pressure exerted by dissolved ions. When the Debye 
screening length $\lambda_D$ is comparable to the colloid size, this pressure 
varies significantly near the particle, with a large magnitude near one end and 
a small magnitude at the other end. In the present experiments, we have 
suspended $0.1$ wt\% of the bigger colloids, where image analysis is more 
accurate. The orientation angle is given in Eq. (\ref{EQ:AR}). As in the 
previous section, the error in the AR measurement is $5$\% because, 
experimentally, it is difficult to focus all colloids in a single frame, and the 
inevitable blur results in inaccuracy in the value of $\text{AR}$. The distribution of 
$\text{AR}$ is obtained by analyzing a single frame consisting of many colloids ($\sim 
10,000$) using the software ImageJ. 

\subsubsection{Electrostatic energy of a colloid near a charged surface}

We calculated the electrostatic energy of a colloid near a charged surface, 
$U_{\rm es}$, by modeling the colloid as a prolate ellipsoid with a semi-major 
axis $a$ and semi-minor axes $b$ and $c<b$. The electrode, located at $y=-d$, is 
covered by a thin dielectric insulating tape with thickness $d$ and permittivity 
$\veps_d$. The interface between 
dielectric and solvent is at $y=0$, and the solvent, with permittivity 
$\veps_s$, is found at $y>0$. 

{\bf Effective potential in the solvent near an electrode covered with a 
dielectric.} In the experiments, the voltage $V$ is applied on the metallic 
electrode, but the voltage the solvent ``feels'', $V_s$, is considerably 
smaller. Once $V_s$ is known, the distribution of the  electric field near the 
colloid allows the calculation of forces acting on the colloid (see below). To 
find $V_s$, we first solved the Laplace and Poisson equations analytically {\it 
in the absence of colloid} as follows. We call $\psi_d(y)$ and $\psi_s(y)$ the 
electrostatic potentials in the dielectric and in the solvent, respectively. 
When ion densities are not too high, the potential $\psi_s$ obeys the 
Poisson-Boltzmann equation
\begin{eqnarray}
\nabla^2\psi_s=\frac{en_0}{\veps_s\veps_0}\sinh\left(\frac{e\psi_s}
{k_BT}\right).\label{PB}
\end{eqnarray}
Here $n_0$ is the bulk ion density far from the charged wall and colloid, 
$\veps_0$ is the vacuum permittivity, $e$ is the electron's charge, $k_B$ is the 
Boltzmann's constant, and $T$ is the absolute temperature. For high ion 
densities, such as in highly charged solutes or surfaces (e.g., in Langmuir 
monolayers), Eq.(\ref{PB}) can be modified by considering the finite size of 
the ions (for more details, see Refs. \cite{ringe2017first,andelman_prl1997}).
The solution of this nonlinear equation in the 
semi-infinite region $y\geq 0$ is 
\begin{eqnarray}
\frac{e\psi_s}{k_BT}=2\ln\left(\frac{1+Be^{-y/\lambda_D}}{1-Be^{-y/\lambda_D}}
\right),
\end{eqnarray}
with the Debye screening length $\lambda_D$ defined in
\begin{eqnarray}\label{eq_debye}
\lambda_D^2=\frac{\veps_s\veps_0 k_BT}{2n_0e^2}.
\end{eqnarray}
The variable $B$ is given by 
\begin{eqnarray}
B&=&\frac{e^{\frac{eV_s}{2k_BT}}-1}{e^{\frac{eV_s}{2k_BT}}+1},
\end{eqnarray}
where $V_s=\psi_s(y=0)$ depends on the experimental potential and needs to be 
found. Note that $\psi_s(y\to\infty)=0$ is obeyed. The electric field in the 
solvent is
\begin{eqnarray}
-\frac{e\psi_s'(y)}{k_BT}=-\frac{1}{\lambda_D}\frac{4B 
e^{y/\lambda}}{B^2-e^{2y/\lambda_D}}.
\end{eqnarray}
In the dielectric, the potential varies linearly, so that 
\begin{eqnarray}
\frac{e\psi_d(y)}{k_BT}=-\frac{eE_d}{k_BT}y+c,
\end{eqnarray}
where $E_d=-\psi_d'$ is the field in the dielectric region.  
There are three variables $c$, $E_d$, and $V_s$ to be found and three 
conditions: (i) the continuity of potential across the dielectric-solvent 
interface at $y=0$, the continuity of displacement field $\veps E$ across the 
interface, and (iii) the external potential at the electrode: 
\begin{eqnarray}
\psi_d(y=0)&=&\psi_s(y=0)\\
-\veps_d\psi_d'(y=0)&=&-\veps_s\psi_s'(y=0)\nn\\
\psi_d(y=-d)&=&V\nn
\end{eqnarray}
From the continuity of the potential at $y=0$, one obtains $c=eV_s/kT$. The 
continuity of the displacement field implies that
\begin{eqnarray}
\veps_d\frac{eE_d}{kT}&=&-\frac{\veps_s}{\lambda_D}\frac{4B}{B^2-1}\nn
\end{eqnarray}
and hence
\begin{eqnarray}
\frac{e\psi_d}{kT}=\frac{\veps_s}{\veps_d}\frac{1}{\lambda_D}\frac{4B}{ 
B^2-1}y+\frac{eV_s}{kT}.
\end{eqnarray}
We finally go back to the equation for the total potential drop across the 
liquid and dielectric, that is,
\begin{eqnarray}\label{eq_Vs}
-\frac{\veps_s}{\veps_d}\frac{d}{\lambda_D}\frac{4B}{ 
B^2-1}+\frac{eV_s}{k_BT}=\frac{eV}{k_BT}.
\end{eqnarray}
This is a highly nonlinear equation for $V_s$ as a function of $V$, 
$d/\lambda_D$, and $\veps_s/\veps_d$. 
For sufficiently small external potentials, the Poisson-Boltzmann equation can 
be linearized. In this Debye-H\"{u}ckel limit, one obtains that the ratio 
between the solvent and external potentials is approximately given by
\begin{eqnarray}
\frac{V_s}{V}\approx \frac{\veps_d\lambda_D}{\veps_s d},
\end{eqnarray}
and this ratio is small since $\lambda_D/d$ is small.

Once $V_s$ is known, whether in the nonlinear or linear regimes, it can be used 
as the boundary condition for the actual solution of the electric field {\it in the 
presence of a colloid} (below). 

{\bf Electrostatic forces acting on the colloid.} 
The electric field is given by the Laplace and Poisson equations in 
the dielectric colloid and in the solvent, respectively: 
\begin{eqnarray}
\veps_c\tilde{\nabla}^2\tilde{\psi}&=&0~~~~~~~~~~~~~~{\rm 
inside~the~colloid}\label{eq_pb}\\
\veps_s\tilde{\nabla}^2\tilde{\psi}&=&\sinh(\tilde{\psi})~~~~~~{\rm 
outside~of~the~colloid}\nonumber
\end{eqnarray}
The potential $\tilde{\psi}=e\psi/k_BT$ and lengths $\tilde{{\bf  
r}}={\bf r}/\lambda_0$ are now written as dimensionless quantities, where 
$\lambda_0$ is given by $\lambda_0^2=\veps_0 k_BT/2n_0e^2$.
Here $\veps_c$ and $\veps_s$ are the colloid and solvent 
permittivities, respectively. The boundary conditions for the potential are 
$\tilde{\psi}(y=0)=eV_s/k_BT$ with $V_s$ found from Eq. (\ref{eq_Vs}) 
and $\tilde{\psi}(y\to\infty)=0$. 

Once the electric field distribution is known, the forces on the colloid are 
found from the Maxwell stress tensor ${\bf T}$
\cite{panofsky_phillips_book}:
\begin{eqnarray}\label{stress_tensor}
{\bf T}_{ij}&=&-p_0(n^\pm,T)\delta_{ij}+\frac12\veps_0\veps 
E^2\left(-1+\rho(\partial\veps/\partial\rho)_T/\veps\right)\delta_{ij}\\
&+&\veps_0\veps E_iE_j.\nn
\end{eqnarray}
$p_0$ is the ideal-gas pressure of the ions: $p_0=(n^++n^-)k_BT$. 
The second term, depending on the colloid's density $\rho$, includes 
electrostriction. It can be lumped together with $p_0$ without changing the 
forces on the colloid. The body force in the liquid ${\bf f}$, given 
as a divergence of the stress, and the force acting on a unit area of the 
interface between the colloid and the solvent, ${\bf f}_s$, are
\begin{eqnarray}\label{body_surface_force}
{\bf f}&=&-\nabla p_0+\frac12\nabla\left(\veps_0E^2 
\rho\frac{\partial \veps}{\partial 
\rho}\right)_T-\frac12\veps_0E^2\nabla\veps
+e(n^+-n^-){\bf E}~,\\
{\bf f}_s&=&{\bf T}\hat{n},
\end{eqnarray}
where $\hat{n}$ is a unit vector perpendicular to the surface of the colloid. 

\begin{figure}[!th]
\begin{center}
\includegraphics[width=0.55\textwidth,bb=0 0 300 
300,clip]{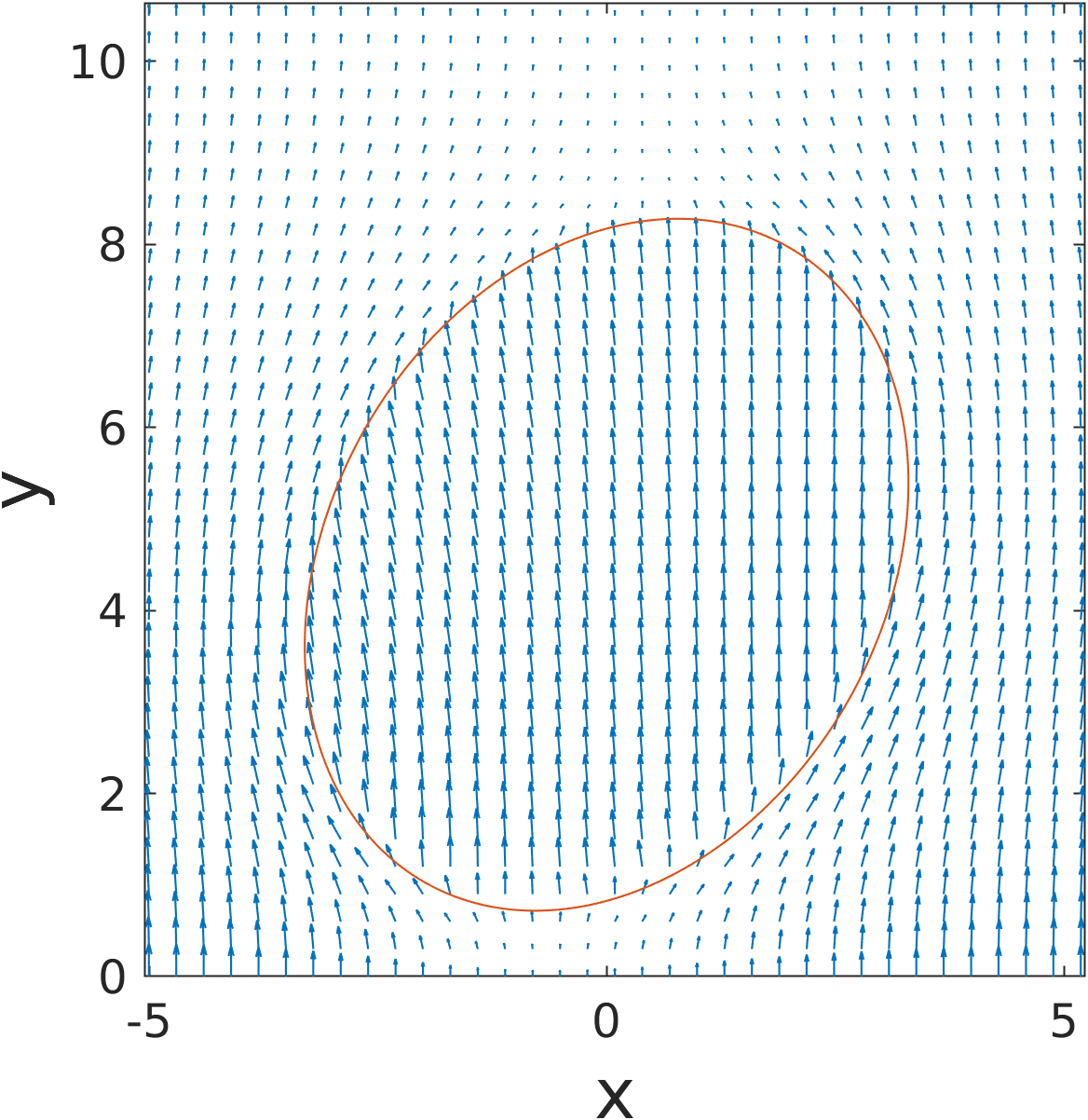}
\caption{Representative electric field lines inside and outside of the 
colloid calculated numerically. The colloid is modeled as an 
ellipsoid (red line), tilted at an angle of $\theta=30^\circ$ with respect to 
the surface normal ($y$-direction). The charged surface is at $y=0$ and the 
solvent is at $y>0$ only. The field lines are the result of a numerical 
solution 
of Eqs. (\ref{eq_pb}) in two spatial dimensions $x$ and $y$. Lengths in $x$ and 
$y$ are scaled by $\lambda_0$ (see text). In this and in other figures we used 
$\veps_c=2.1$, $\veps_s=48$, and $a=4\lambda_0$, and $a/b=1.32$.
}
\label{field_lines}
\end{center}
\end{figure}

In the above formalism, one finds the electrostatic potential $\tilde{\psi}$ 
and surface force ${\bf f}_s$, then integrates the total torque 
$\tau(\theta)=\int_s{\bf r}\times{\bf f}_s ds$ over all elements of 
the colloid's surface $ds$. When the torque $\tau(\theta)$ is known, the 
effective electrostatic rotational potential $U_{\rm es}(\theta)$, is defined 
as 
\begin{eqnarray}\label{Ues_def}
\tau(\theta)=-\frac{U_{\rm es}(\theta))}{d\theta}.
\end{eqnarray}

DMSO is used as a solvent because of its high dielectric constant 
$\veps_s\approx 48$ and extremely low density of charge carriers. DMSO is a 
weak acid and, in contrast to water, the charges on the molecules are induced 
and not independent. From the density $\rho=1.1$ g/cm$^3$ and molar mass 
$m=78.13$ g/mol, one obtains the volume of DMSO molecule as $v_0\approx  
1.1\times 10^{-22}$ cm$^3$. The known pKa value of $35$ allows to calculate the 
fraction $f$ of charged molecules as $f\sim 10^{-18}$, which is well below the 
detection level of most measurement methods. The small residual ionic 
conductivity in DMSO is due to the unknown amount of impurities and leads to a 
Debye screening length $\lambda_D$ of the order of one $\mu$m. The fact that
the particle size is comparable to $\lambda_D$ is meaningful for the physics of 
particle orientation -- if an elongated particle is too small compared to 
$\lambda_D$, it ``feels'' a field that is effectively uniform, and it will 
orient parallel to the field. If the particle is too large compared to 
$\lambda_D$, the field decays too strongly near the surface and the particle 
does not feel significant forces. 

\begin{figure}[!th]
\begin{center}
\includegraphics[width=0.65\textwidth,bb=0 0 380 
300,clip]{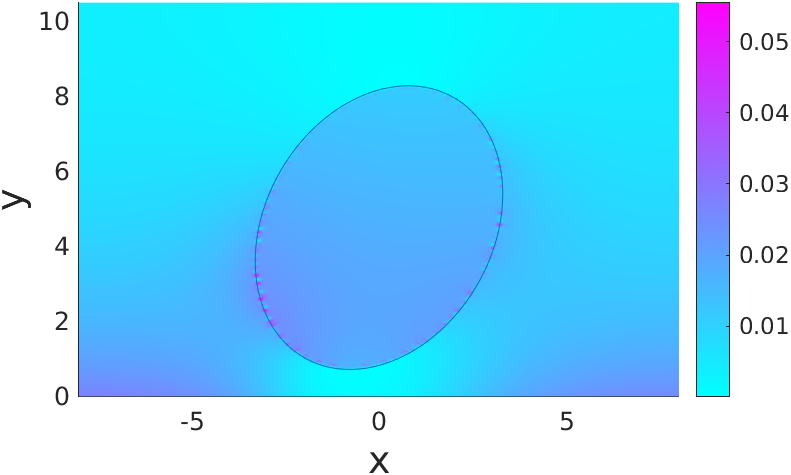}
\caption{A plot of $E^2(x,y)$ for the same colloid as in Fig. 
\ref{field_lines}. Color bar expresses the intensity of field squared in 
dimensionless units. Lengths in $x$ and $y$ are scaled by $\lambda_0$. Here the 
surface potential is given by $eV_s/k_BT=1$.
}
\label{field_intensity}
\end{center}
\end{figure}

Figure \ref{field_lines} shows a representative numerical calculation of the 
field distribution in two spatial dimensions around the colloid. Field lines 
are derived from the potential $\tilde{\psi}$ that we found from Eqs. 
(\ref{eq_pb}). The surface at $y=0$ represents the dielectric-solvent interface, 
and its potential is $\tilde{\psi}=\tilde{V}_s=eV_s/k_BT$, with $V_s$ taken as 
the solution of Eq. (\ref{eq_Vs}) with a particular value of electrode potential 
$V$. All lengths are scaled by $\lambda_0$. The field lines inside the colloid 
are nearly uniform; one may recall that the field lines inside a perfect 
dielectric ellipsoid are uniform if the external field is uniform. Here the 
external field is inherently nonuniform, resulting in deviations near the 
boundary of the colloid (red contour line).

The electrostatic force in Eq. (\ref{body_surface_force}) depends quadratically 
on the field. Figure \ref{field_intensity} shows the distribution of field 
intensity $E^2(x,y)$ inside and near the same colloid as in Fig. 
\ref{field_lines}. The field's intensity is somewhat diminished between the 
colloid and the surface ($x\approx 0$, $0\leq y\leq 1$) and sufficiently far 
from the surface ($y\geq 6$). In these two figures, the tilt angle is 
$\theta=30^\circ$; to obtain $U_{\rm es}(\theta)$, we repeated the 
calculation for a series of $\theta$ values in the range $0\leq\theta\leq 
90^\circ$.

\subsubsection{Gravitational energy of the colloid}

In addition to the electrostatic energy of the colloid, we needed to add the 
gravitational energy of a tilted ellipsoid. The height of the center of mass of 
the ellipsoid above the surface as a function of the tilt angle $\theta$ 
of the long axis with respect to the normal to the surface, $h(\theta)$, is
\begin{equation}
h(\theta)=\sqrt{b^2\sin^2\theta+a^2\cos^2\theta}.
\end{equation}
When $\theta=0$, the colloid is normal to the surface and the 
height of the center of mass is $h=a$; when the ellipsoid is lying parallel to 
the surface, $\theta=90^\circ$ and $h=b$. The effective energy for the 
displacement of the center of mass of the ellipsoid is $U_g=V\Delta\rho g
h$, where $V=(4\pi/3)abc$ is the ellipsoid's volume, $\Delta\rho$ is the 
density difference between the colloid and the surrounding solvent, and $g$ is 
the gravitational acceleration. We thus find 
\begin{equation}\label{Ug}
U_g(\theta)=\frac{4\pi}{3}abc\Delta \rho g 
\sqrt{b^2\sin^2\theta+a^2\cos^2\theta}.
\end{equation}
This energy is minimal when the colloid lies with its long axis parallel to the 
surface. It is instructive to estimate the order of magnitude of $U_g$. For a 
$\mu$m-sized colloid with $a=b=c=4$ $\mu$m and $\Delta\rho=1$ kg/m$^3$, we have 
$U_g\approx k_BT$, and thus for colloids of similar size the gravitational 
energy is not negligible compared to the thermal energy. 
\begin{figure}[!th]
\begin{center}
\includegraphics[width=0.6\textwidth,bb=0 0 450 
500,clip]{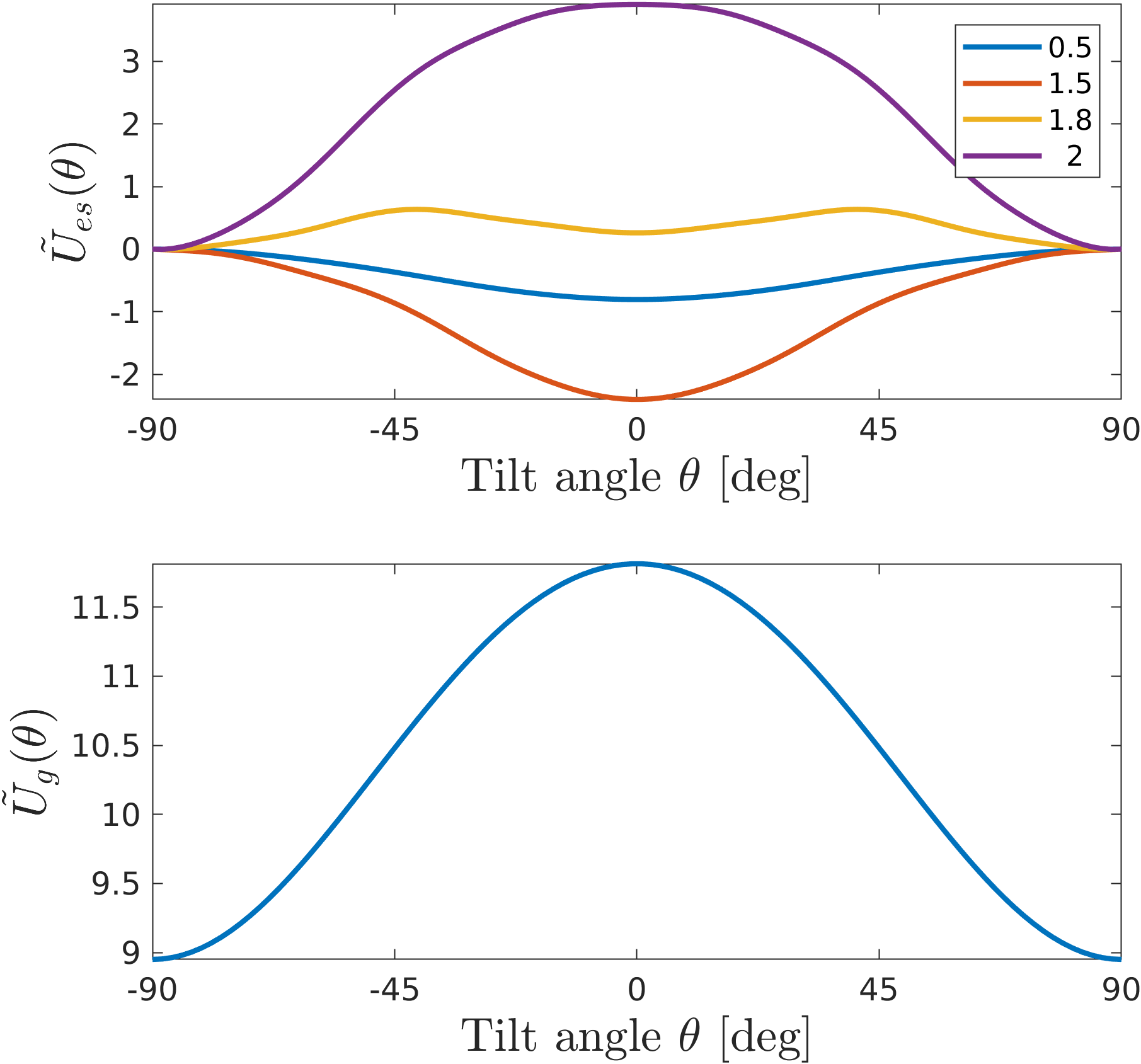}
\caption{Top: Colloid's electrostatic energy as a function of its tilt 
angle $\theta$, $\tilde{U}_{\rm es}=U_{\rm es}/k_BT$ with $U_{\rm es}$ defined 
in Eq. (\ref{Ues_def}), for different external potentials. Legend shows the 
scaled surface potential $eV_s/k_BT$. Bottom: Scaled colloid potential 
energy due to gravity vs. $\theta$: $\tilde{U}_g=U_g/k_BT$ with $U_g$ from Eq. 
(\ref{Ug}). We used $b=5.1~\mu$m, $a=6.7~\mu$m, $c=b$, $\Delta\rho=1$ kg/m$^3$, 
and the ratio between the length in the $z$ direction and $l_{B0}$ is $5$, 
where 
$l_{B0}=e^2/(4\pi\veps_0 k_BT)$.}
\label{Ues_Ug}
\end{center}
\end{figure}

Figure \ref{Ues_Ug} shows the dimensionless electrostatic $\tilde{U}_{\rm 
es}=U_{\rm es}/k_BT$ (top) and 
gravitational $\tilde{U}_g=U_g/k_BT$ (bottom) energies vs. colloid tilt angle. 
At small surface potentials (value in legend), $\tilde{U}_{\rm es}$ has a 
global minimum at $\theta=0$, indicating that colloids tend to orient 
parallel to the field and normal to the surface. When the potential increases, 
the minimum becomes deeper. However, due to the ``ideal gas'' pressure of ions 
and its nonuniform distribution near the surface, this behavior is 
non-monotonous -- if the potential increases further, the global minimum 
becomes local, and the global minima appear at $\theta=\pm 90^\circ$ (both 
orientations are equivalent), indicating that the most stable colloid 
orientation is parallel to the surface. As the voltage increases, the local 
minimum at $\theta=0$ completely disappears. The dependence of $\tilde{U}_g$ is 
independent of voltage and its minimum is always at $\theta=\pm 90^\circ$. 
\begin{figure}[!th]
\begin{center}
\includegraphics[width=0.6\textwidth,bb=0 0 400 
400,clip]{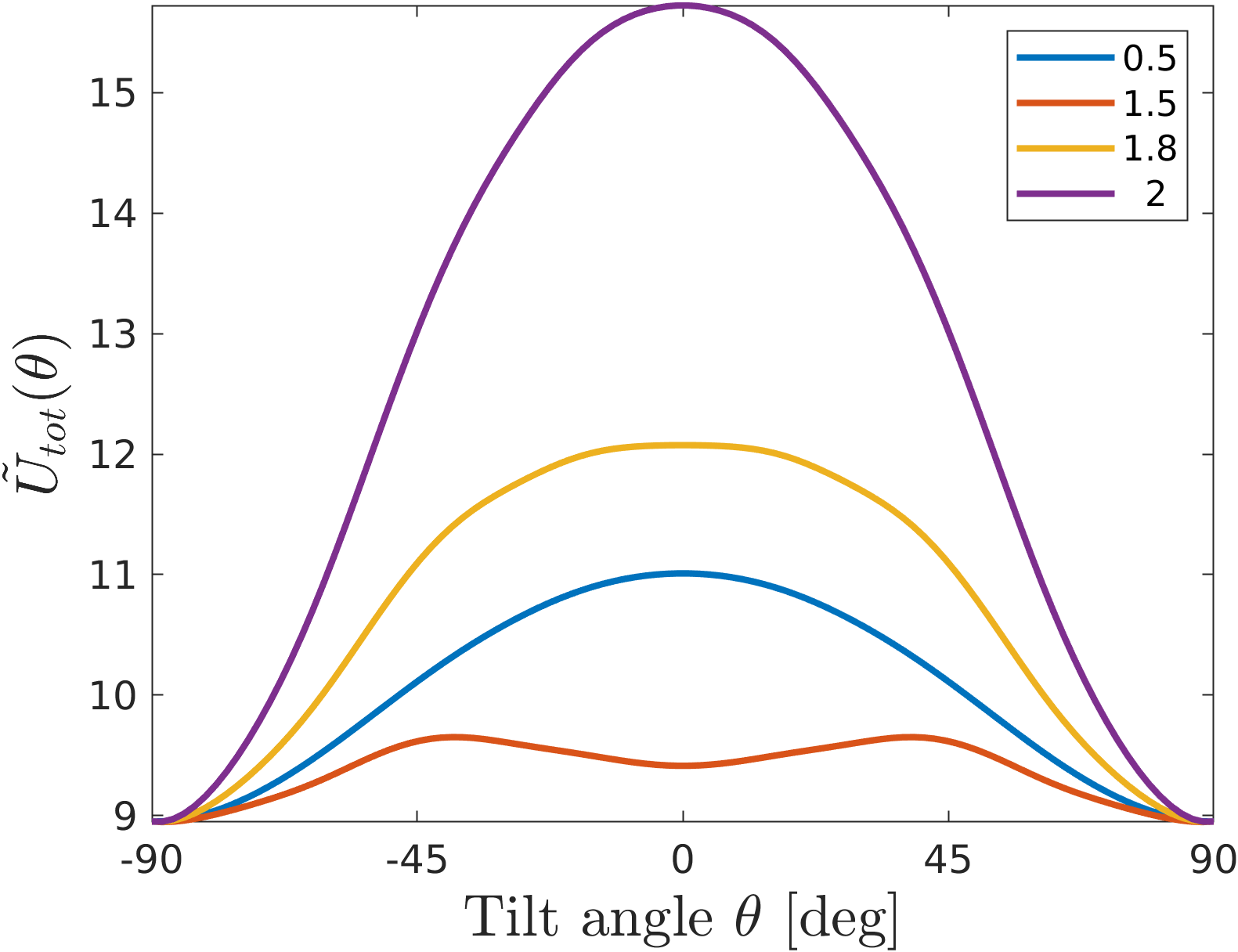}
\caption{The sum of electrostatic and gravitational energies, $\tilde{U}_{\rm 
tot}=\tilde{U}_{\rm es}+\tilde{U}_g$ from Fig. \ref{Ues_Ug} for different 
surface potentials $\tilde{V}_s=eV_s/k_BT$ (legend). 
}
\label{Utot}
\end{center}
\end{figure}

The sum of electrostatic and gravitational contributions to the energy, 
$\tilde{U}_{\rm tot}=\tilde{U}_{\rm es}+\tilde{U}_g$, is plotted in Fig. 
\ref{Utot}. At small values of 
$\tilde{V}_s$, the gravitational part dominates the energy and the global 
minima are at $\theta=\pm 90^\circ$. As the potential increases, the 
electrostatic minimum is large enough and the global minimum 
shifts to $\theta=0$ (colloid normal to the surface). Further increase of 
$\tilde{V}_s$ leads to re-stabilization of the minima at $\theta=\pm 90^\circ$. 
\begin{figure}[ht!]
\begin{minipage}{1\textwidth}
\begin{center}
\includegraphics[width=0.6\textwidth,bb=0 0 450 
400,clip]{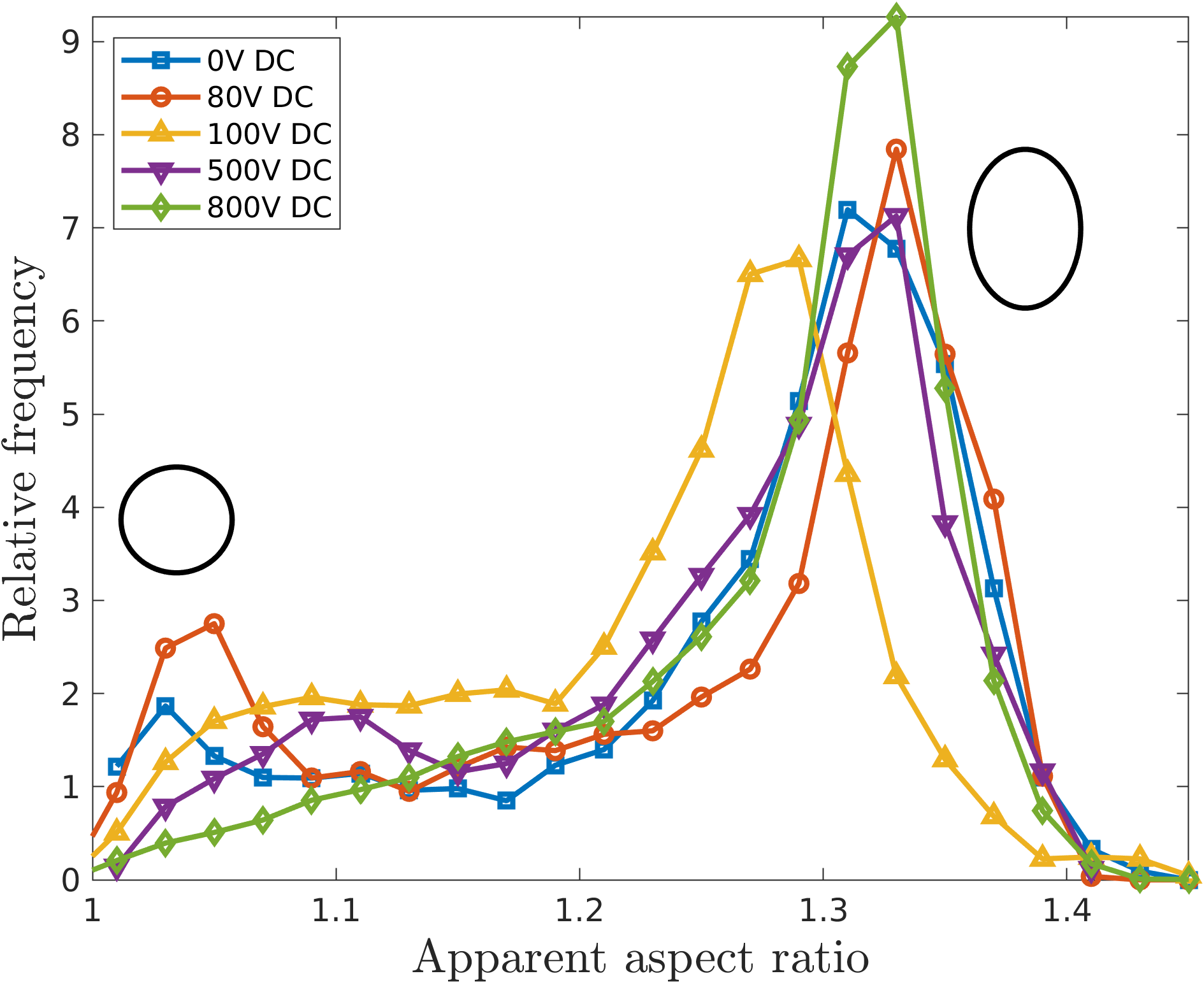}
~\\~\\~\\
\includegraphics[width=0.6\textwidth,bb=0 0 450 
400,clip]{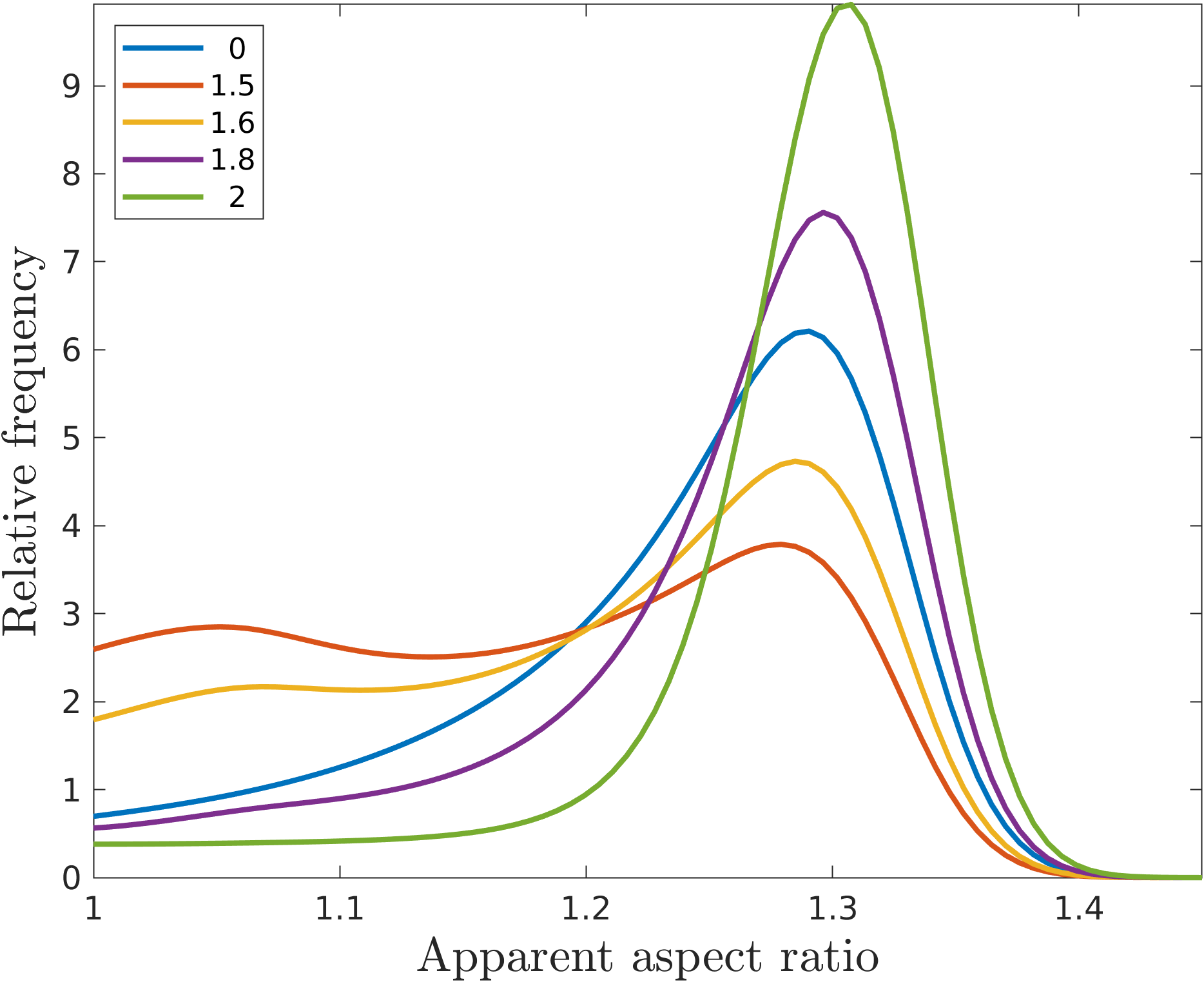}
\end{center}
\end{minipage}
\caption{Top: Apparent aspect ratio $\text{AR}$ distribution for colloids near the 
charged surface in DC fields (electrode potential $V$ is indicated in 
the legend). The relative frequency is calculated by counting all the 
particles with a certain aspect ratio and dividing by the total number of 
particles ($\sim 10,000$). Circle and ellipse illustrations indicate vertical 
and horizontal particle orientations, respectively. Bottom: Theoretical results 
corresponding to the top panel. Legend shows the dimensionless surface potential 
$\tilde{V}_s$ calculated by Eq. (\ref{eq_Vs}) for a given value of $V$. We used 
$a=4\lambda_0$, and the center-of-mass distance of the colloid from the $y=0$ 
surface is $4.5\lambda_0$. }
\label{fig_rel_freq}
\end{figure}

In the experiments, the colloids are ``polydisperse'', with an average value of 
the aspect ratio $a/b$ and a finite probability for larger or smaller values. To 
account for this, we assumed that $a$ obeys a normal distribution, that is, the 
probability of having $a$ is given by $P(a)=\exp(-(a-a_{\rm av})^2/2a_{\rm 
sd}^2)/\sqrt{2\pi a_{\rm sd}^2}$, where $a_{\rm av}$ is the average value and 
$a_{\rm sd}$ is the standard deviation. We made extensive numerical calculations 
of $U(\theta)=U_{\rm es}(\theta)+U_g(\theta)$ for all values of $\theta$ and for 
three values of $a$: $a=1.27b$, $a=1.32b$, and $a=1.37b$. For a given value of 
$a$, the apparent aspect ratio $\text{AR}$ as a function of $\theta$ is given by Eq. 
(\ref{EQ:AR}). The relative frequency of colloids at angle $\theta$ associated 
with this value of $a$ is proportional to the Boltzmann weight 
$\exp(-U(\theta))$. We then calculated the value of $U(\theta)$ for {\it all} 
values of $a$ by interpolating $U(\theta)$ from the three exact values, with 
the correct weight $P(a)$ using $a_{\rm av}=1.32 b$ and $a_{\rm sd}=0.03 b$. 
The total relative frequency of occurrence of $\text{AR}$ is the sum of frequencies 
from all possible values of $a$.

The experiments and theory are displayed in Fig. \ref{fig_rel_freq}. The top 
panel shows the relative frequency of the apparent aspect ratio as observed 
experimentally in a sample of $\sim 10,000$ particles. Different curves relate 
to the different voltages indicated in the legend. The $y$-axis is normalized 
such that the integral of each curve is unity. For all voltages, the maxima of 
the distribution are found at $AR\sim 1.32$, which corresponds to the ratio 
$a/c$ of the ellipsoid, indicating particles lying flat at the surface. A 
smaller peak appears at small voltages around $AR=1.05$. This corresponds to the 
ratio $b/c$ of the ellipsoid, indicating colloids oriented normal to the surface 
(parallel to the field). The dependence of this peak on potential is 
non-monotonic -- the peak height increases with increasing potentials and then 
decreases. The measurements are in equilibrium -- an increase or decrease of 
the voltage followed by a return to the original value always leads to the same 
particle distribution. The theoretical results (bottom panel) show the same 
behavior semi-quantitatively. In particular, the theory predicts a large peak at 
$AR\approx 
1.32$ and a smaller peak at $AR\approx 1.05$, which disappears at large 
voltages.

\section{Conclusions} 

We studied the bistability of elongated ellipsoidal particles in time-varying 
AC and static DC electric fields. Two parallel ITO glass slide electrodes 
produced the field. The external frequency is a convenient handle that allows 
tuning the range of the interaction exerted by the charged surface (electrode) 
on the particles; the potential determines the amplitude of the interaction. In 
the first part, we used AC fields at the above-kHz frequency, where the field 
is effectively spatially uniform. This is the limit where the influence of the 
surface has an infinite range. The extensive mapping of the large 
potential--frequency parameter space is absent from previous works. The phase 
diagram in the potential--frequency plane is divided by a line into two regions. 
At low surface potentials, particles lay parallel to the surface. An increase of 
the frequency above the threshold line leads to their alignment normal to the 
surface. At large surface potential, the same transition happens but at a 
smaller threshold frequency. For larger particles, this line is displaced to 
larger potentials and frequencies (see Fig. \ref{fig_phase_diagram}).

While in AC fields, the force acting on the particles is spatially uniform, in 
DC fields, the physical picture is very different: here, screening occurs, and 
the field is substantial only sufficiently close to the electrode. Thus, the 
forces acting on a particle can be drastically different on the side near the 
electrode compared to the remote side. Because the particles are nonspherical, 
they feel a torque. Theory shows that this torque depends on the potential, 
location, tilt angle, and particle shape (given mainly by the ratio $a/b$ of the 
two ellipsoidal axes). We extended the previous theory to include the effect of 
gravity, coupling to the difference between the specific densities of particles 
and solvent, and favoring particles lying flat on the surface. We verified 
experimentally the theoretical prediction that particle orientation is bistable, 
with a discontinuous first-order transition between particle orientation 
parallel to the surface to normal orientation. 

The current study may have applications in various areas, such as in the 
fabrication of 3D non-corrosive and lower band-gaps with diamond-like lattice 
photonic crystals from anisotropic particles. It may help mimicking atoms and 
molecules with diverse geometry, chemical composition, and surface 
functionalities. Another interesting direction is related to plasmonics -- it is 
well known that plasmons have different wavelengths parallel to the long or 
short axis of elongated particles. The switching behavior discovered here may 
allow for tuning the plasmon frequency of a colloidal array on a surface between 
two states. The present study advances the fundamental understanding of the	
forces acting on particles in electrolytes in electric fields. It serves as a 
first step toward a study of more complex phenomena that occur when such 
particles pack in high densities on a surface, where steric interactions are 
dominant.

It would be interesting to extend the current work to frequencies lower 
than the kHz range. At these frequencies, the distance 
dissolved ions traverse during one field cycle can be large, and Joule heating 
can be substantial. There is a fundamental question on the effect of this 
heating on the phenomena we observed here, especially recalling that heating 
would be more pronounced near the surface compared to far from it. Another, more 
application-oriented direction, is the possibility of controlling the optical 
properties of a surface: if the colloidal density is sufficiently large, we 
speculate that radiation impinging on the liquid-crystalline-like surface could 
be scattered, reflected, or polarized, depending on the orientation of the 
colloids.

\section*{Acknowledgement}

This work was supported by the Israel Science Foundation (ISF) grant No. 274/19.



\end{document}